\documentclass[%
 reprint,
 amsmath,amssymb,
 aip,
 pop, 
]{revtex4-1}

\usepackage{graphicx}
\usepackage[font=small,labelfont=bf,
   justification=justified,
   format=plain]{caption}
\usepackage{dcolumn}
\usepackage{bm}
\usepackage{multirow}
\usepackage{subfigure} 
\usepackage{amsmath}
\usepackage{subcaption}
\usepackage{aas_macro}
\usepackage{xcolor}
\usepackage[table]{xcolor}

\begin{document}

\title{Radiative PIC simulations of relativistic pair plasma: multiple interacting current sheets and turbulent evolution.}

\author{Fulvia Pucci}
\email{fulvia.pucci@inaf.it}
\affiliation{INAF, Osservatorio Astrofisico di Arcetri, Largo E. Fermi 5, I-50125 Firenze, Italy}
\affiliation{SETI Institute, Mountain View, CA, USA}

\author{E. Amato}
\affiliation{INAF, Osservatorio Astrofisico di Arcetri, Largo E. Fermi 5, I-50125 Firenze, Italy}
\affiliation{Dipartimento di Fisica e Astronomia, Università degli Studi di Firenze, Via G. Sansone 1, I-50019 Sesto F. no (Firenze), Italy}

\author{Dario Borgogno}
\affiliation{ IAPS, INAF, Via del Fosso del Cavaliere, 100, 00133 Roma RM, Italia}

\author{N. Bucciantini}
\affiliation{INAF, Osservatorio Astrofisico di Arcetri, Largo E. Fermi 5, I-50125 Firenze, Italy}
\affiliation{Dipartimento di Fisica e Astronomia, Università degli Studi di Firenze, Via G. Sansone 1, I-50019 Sesto F. no (Firenze), Italy}
\affiliation{INFN, Via G. Sansone 1, I-50019 Sesto F. no (Firenze), Italy}

\author{P. Buratti}
\affiliation{ IAPS, INAF, Via del Fosso del Cavaliere, 100, 00133 Roma RM, Italia}
\affiliation{ENEA, NUC Department, Via E. Fermi 45, 00044 Frascati, Italy}

\author{M. E.  Innocenti}
\affiliation{ Institut für Theoretische Physik, Ruhr-Universität Bochum, Bochum, Germany}

\author{K. M. Schoeffler}
\affiliation{ Institut für Theoretische Physik, Ruhr-Universität Bochum, Bochum, Germany}

\author{Marco Tavani}
\affiliation{ IAPS, INAF, Via del Fosso del Cavaliere, 100, 00133 Roma RM, Italia}
\date{\today}

\author{Valerio Vittorini}
\affiliation{ IAPS, INAF, Via del Fosso del Cavaliere, 100, 00133 Roma RM, Italia}

\begin{abstract}
Two-dimensional relativistic particle-in-cell (PIC) simulations of radiative magnetic reconnection in pair plasmas with multiple interacting current sheets are carried out to mimic the dynamics in high-energy astrophysical environments, such as particle acceleration regions in pulsar wind nebulae and relativistic outflows, where the magnetic field is expected to reverse polarity multiple times. Initially, due to reconnection within each isolated sheet, particles are accelerated and synchrotron emission beyond the burn-off limit is confirmed, even if the particle distribution function shows steep slopes. After this phase, plasmoids lead to cross-sheet interactions and merging, with new current sheets formed. In this regime  a Kolmogorov-like spectrum for the magnetic energy develops over a couple of decades, followed by a dissipation range starting around 5~$d_e$ (electron inertial lengths), showing that multi-sheet reconnection evolves nonlinearly into well-developed turbulence. 
This phase provides secondary acceleration and further cooling by synchrotron emission, with intermittent radiative bursts. We show that high energy accelerated particles by the primary current sheets are further energized during the turbulent phase, while the distribution of the most energetic particles remains steep. 

\end{abstract}

\maketitle

\section{Introduction}
Magnetic reconnection is widely recognized as a key mechanism for the rapid conversion of magnetic energy into plasma heating, bulk flows, and non-thermal particle acceleration across astrophysical and laboratory environments. 
Recently, systems exhibiting multiple current sheets have become a focus of growing interest.
In the heliosphere, the heliospheric current sheet (HCS; \citep{Smith2001}) is typically observed to be stable in the solar wind at 1 AU; yet, in the inner heliosphere, signatures of magnetic reconnection in the near-Sun HCS are observed with surprising frequency by Parker Solar Probe~\citep{phan2021prevalence, phan2022parker, du2025occurrence}. The HCS can become compressed and unstable near the heliospheric termination shock \citep{2011ApJ...734...71O}. Spacecraft observations in the solar wind often reveal that single-sector boundaries contain multiple subscale current layers (i.e., thinner than the HCS), interpreted and modeled as the folding or interaction of individual magnetic flux tubes \citep{1993JGR....98.9371C, Maiewski2020, phan2024multiple}.
In the magnetosphere, localized double layers and intense parallel electric fields can transiently appear within and between current layers, mediating fast particle acceleration and the generation of flat-top or non-thermal distributions \citep{Egedal2015, Singh2011}. 
The coupling between reconnection and the emergence of such structures is now recognized as a distinctive feature of kinetic-scale turbulence. In the case of finite plasma $\beta_p$ (the ratio of the plasma pressure to the magnetic pressure) island growth and related particle acceleration has been investigated with particle-in-cell simulations \citep{2010ApJ...709..963D, Schoeffler_2011, Schoeffler_2012,Schoeffler_2013}.
\subsection{Multi-layered Structures in Pulsar Winds}
Similar multi-layer structures naturally arise in pulsar environments.
Here, the rapid rotation and strong magnetic field of a neutron star induce huge electric fields that pull charges off the surface, setting up currents and accelerating particles to energies high enough to emit gamma rays. At these energies, photon–photon or photon–magnetic-field interactions then create cascades of electron–positron pairs that fill the magnetosphere. The alternating magnetic polarity of the rotating magnetosphere induces alternating magnetic field polarity in the subsequent pulsar wind — the “stripes”—separated by current sheets dominated by the electron–positron pairs \citep{Coroniti90,Lyuba2001,Lyubarsky2005,2018ApJ...855...94P,2019ApJ...876L...6P, Cerutti20,2020A&A...642A.204C}. 
Magnetic reconnection in pair plasmas is central to understanding how striped pulsar winds convert their enormous Poynting flux into the energetic particles and radiation. The dissipation of these sheets — through reconnection and subsequent plasmoid formation — is proposed as the key mechanism enabling the conversion of Poynting-dominated outflows into particle-dominated winds and high-energy radiation.
Particle-in-cell simulations have shown that when these current sheets interact with the pulsar termination shock, they can trigger efficient particle acceleration and strong synchrotron emission \citep{Nagata2008, Sironi2011, Cerutti2020, Lu2021}. 
\subsection{PIC simulations of multiple current sheets dynamics to explain Flares in the Crab Nebula.}
Recent kinetic studies have demonstrated that when the reconnection outflows and secondary current sheets coexist, the system naturally evolves toward a turbulent, multi-scale state characterized by magnetic island coalescence (e.g., \citealt{Sironi2011,Cerutti2013}). 
In the relativistic regime, particle-in-cell (PIC) simulations with synchrotron or inverse-Compton radiation reaction have revealed that radiative feedback can drastically modify reconnection dynamics. Early works by \citep{Cerutti2013, Cerutti2014, Sironi2019}
 showed that strong cooling compresses current sheets, enhances field intensities, and produces ultra-compact, radiatively efficient plasmoids—potentially explaining fast $\gamma$-ray flares from systems such as the Crab Nebula and black hole coronae. More recently, \cite{Mehlhaff2020, Schoeffler2023} by means of 2D and 3D radiative reconnection simulations, demonstrated how synchrotron back reaction regulates plasmoid compression and introduces intermittent, beamed radiation patterns. These studies collectively indicate that radiative cooling is not a passive energy sink, but an active agent shaping both field topology and particle dynamics.
Despite these advances, multi-layer configurations—where multiple current sheets form, interact, and merge—remain largely unexplored under radiative conditions. Yet, such configurations are expected in global systems where reconnection drives secondary tearing and cascading instabilities, ultimately producing a turbulent hierarchy of current structures. Exploring these processes in a controlled setup is thus essential for connecting kinetic-scale dissipation to macroscopic variability in high-energy sources.
\subsection{This work.}
In this work, we present two-dimensional PIC simulations performed with \textsc{Zeltron} \citep{Cerutti2013}, designed to investigate the radiative evolution of multiple interacting current sheets. Our configuration extends the setup discussed in \citealt{Cerutti2013}, i.e the Crab flaring emission region, in the context of a more realistic multiple current sheet configuration, following the evolution to a nonlinear later stage. Indeed, we expect that within the Crab flaring emission region considered by \citealt{Cerutti2013} -- with size estimated to be $L \sim 7\times 10^{15}$ cm, see e.g.  \citealt{Vittorini_2011} -- multiple current sheet will be present. This work also extends
 \citealt{10.3389/fspas.2022.954040} to include synchrotron back reaction, allowing us to study how the interaction between the adjacent current sheets changes the radiative losses and the latter in turn affect the dynamics, energy partition, and the properties of the electromagnetic fields and accelerated particles. To capture the essential physics of radiative magnetic reconnection in a multi-layer system while retaining full kinetic and radiation feedback, we consider a 2D periodic domain containing four parallel current sheets. This configuration can be regarded as the minimal kinetic realization of a “multiple current sheet” system: it allows for the formation, interaction, and coalescence of plasmoids both within each sheet and between different sheets, while remaining computationally tractable for fully relativistic, radiative PIC simulations. 
By comparing radiative and non-radiative runs, we assess how feedback from synchrotron back reaction modifies the hierarchy of current layers and the resulting particle and radiation spectra. These results provide new insight into environments where magnetic reconnection competes with other -- secondary-- instabilities.

\section{Simulation Setup and Parameters}
We perform two-dimensional relativistic particle-in-cell (PIC) simulations of a pair-plasma using the code \textsc{Zeltron} \citep{Cerutti2013}. 
The domain contains four Harris-type current sheets centered at positions $y_i$ ($i=1\dots4$) so that all the current sheets are at the same distance from the nearest neighbors $y_s=L_y/4$ --where $L_y$ is the length of the box in the y-direction-- and the first current sheet from the lower boundary is located at $y_{offset}=L_y/8$. We chose the Harris equilibrium as it is an exact equilibrium of the Vlasov–Maxwell system: no artificial transient is introduced as might occur with a force free equilibrium (see e.g. \citealt{An_2023}). We use biperiodic boundary conditions. Compared to the extensively studied single- and double-current-sheet setups (e.g.,~\citealt{schoeffler2025particle, schoeffler2025energetic}), a four-layer system introduces new behavior and further plasma acceleration, as we will show here. At the same time, a configuration with four sheets is still small enough to allow us to resolve the relevant kinetic and radiation-reaction scales over many dynamical times.
The in-plane magnetic field is derived from a vector potential oriented along $z$,
\begin{equation}
A_z(y) = B_0 a s_i \ln\!\left[\cosh\!\left(\frac{y - y_i}{a}\right)\right],
\end{equation}
where $B_0$ is the asymptotic upstream magnetic field. All the values in CGS units are shown in Table~\ref{tab:params}.
The half-thickness of each sheet $a$ is defined as\citep{Cerutti2013}
\begin{equation}
a=2\frac{k_B T_d}{\beta_d \gamma_d\,e\,B_0},
\end{equation}
with $\gamma_d= (1-\beta_d^2)^{-1/2}$, where $\beta_d=v_d/c$, the drifting population speed, is also shown in Table~\ref{tab:params}.  The pressure balance is localized at the current sheets, i.e. $a\ll y_s$ then the rest of the domain results to be magnetically dominated.
The equilibrium fields are defined separately over four intervals:  $[0\le y< 1/4]$: $y_1=1/8$, $s_1 =-1$; $[1/4\le y < 1/2]$: $y_2=3/8$, $s_2 =1$; $[1/2\le y< 3/4]$: $y_3=5/8$, $s_3=-1$;
$[3/4\le y< 1]$: $y_4=7/8$,  
$s_4 =1$.
This configuration yields
\begin{equation}
\mathbf{B}(y) = \frac{dA_z}{dy}\,\hat{\mathbf{x}}
= B_0 s_i \tanh\!\left(\frac{y - y_i}{a}\right)\,\hat{\mathbf{x}},
\end{equation}
and an equilibrium current density
\begin{equation}
\mathbf{J}(y) = \,\frac{c B_0}{4\pi a} s_i\,\mathrm{sech}^2\!\left(\frac{y - y_i}{a}\right)\,\hat{\mathbf{z}},
\end{equation}
(CGS units are used unless otherwise stated).
The minimal Larmor radius for the background particles is defined as:
\begin{equation}
\rho_c = \frac{k_B T_d}{e B_0},
\end{equation}
where $T_d$ is the temperature of the drifting electrons.

We allow the tearing instability to develop spontaneously from particle noise as we noticed that introducing perturbations might drive preferential secondary merging between the current sheets, which is one of the primary focus of this work. 

The density of the plasma at the current sheets is obtained using $\mathbf{J}=e(n_p\mathbf{v}_p-n_e\mathbf{v}_e)$, i.e. the current is carried by counter-streaming electrons and positrons moving at $v_s=\pm\beta_d c$, where s =p,e labels the species and $n_p \sim n_e\sim n_0$:
\begin{equation}
2 e n_0 \beta_d c = \frac{c}{4\pi a}B_0,
\end{equation}
which yields the density at the sheet center $n_0 = B_0/(8\pi e \beta_d a)$.

The plasma density profile follows the Harris equilibrium with a uniform background:
\begin{equation}
n(y) = n_b + n_0\,\mathrm{sech}^2\!\left(\frac{y - y_i}{L}\right),
\end{equation}
where $n_0$ and $n_b$ are the drifting and background densities, respectively. The locations $y_i$ corresponding to the density peaks are the same as the current sheet locations, described above, to provide a pressure equilibrium configuration. 

We define the ``hot'' magnetization parameter as 
\begin{equation}
\sigma_{\rm hot} = \frac{B_0^2}{4\pi w},
\end{equation}
whose value is shown in Table~\ref{tab:params}, where $w\simeq n \gamma_1 m c^2$ is the enthalpy of the system, since we assume $\gamma_1\gg1$ so that the thermal contribution dominates the plasma energy content. In terms of the usual definition of the magnetization parameter $\sigma_{\rm cold} $, which compares the magnetic energy density with the rest–mass energy density of the plasma, we have $\sigma_{\rm hot} = \frac{ \sigma_{\rm cold}}{\gamma}$ (see e.g. \citealt{Zenitani2007,Sironi2014}). Time is normalized using the inverse Larmor frequency $\omega_c=\rho_c/c$, where $\rho_c$ is the minimum Larmor radius, listed in Table~\ref{tab:params}.
This setup matches the configuration employed in \citet{Cerutti2013}, for comparisons to our multi-sheet geometry. We also adopt the same initial background particle energy distribution, i.e. a power-law with index $p$ between $\gamma_1$ and $\gamma_2$, see Table~\ref{tab:params}. In \citet{Cerutti2013}, even if the energy associated with the Crab Nebula background is much lower than $\gamma_1$, the choice of the initial energy range is justified by the computational time available. The drifting particle distribution-- supporting the current sheet-- is assumed to be a Maxwell-Jüttner with drift velocity $\beta_d c$ and normalized temperature $T_d$. All the values in CGS are described in Table~\ref{tab:params}.
The goal of this work is to compare the evolution of the two sheet configuration with the case of multiple current sheets and to determine how the development of turbulence affects particle acceleration and synchrotron emission.

\begin{table*}[h]
\caption{\label{tab:params}Physical and numerical parameters adopted in the simulations.}
\begin{ruledtabular}
\begin{tabular}{lc}

\rowcolor{gray!25}\multicolumn{2}{l}{\textbf{Current sheets}} \\
Upstream magnetic field $B_0$ & $5~\mathrm{mG}$ \\
Layer half-thickness $a/\rho_c$ & $2.7$ \\
\hline

\rowcolor{gray!25}\multicolumn{2}{l}{\textbf{Particle distributions}} \\
Drift particles velocity $\beta_{\mathrm{drift}}$ & $0.6$ \\
Drift particles temperature $k_B T/mc^2$ & $4\times 10^7$ \\
Background density $n_{\mathrm{bg}}/n_0$ & $0.1$ \\
Background distribution energy range ($\gamma_1$, $\gamma_2$) & $4\times10^7$, $4\times10^8$ \\
Background distribution power-law index $p$ & $2$ \\
\hline

Radiation-reaction limit $\gamma_{\mathrm{rad}}$ & $1.3 \times 10^{9}$ \\
Minimum Larmor radius $\rho_c$ & $1.4 \times 10^{13}~\mathrm{cm}$ \\
Magnetization $\sigma$ & $16$ \\
\hline

\rowcolor{gray!25}\multicolumn{2}{l}{\textbf{Numerical Parameters}} \\
Resolution $\rho_c/\Delta x = \rho_c/\Delta y$ & $3$ \\
Time step $\Delta t\,\omega_1$ & $0.07$ \\
Particles per cell (4 species) & $100$ \\
Domain size $L_x/\rho_c = L_y/\rho_c$ & $500$ \\

\end{tabular}
\end{ruledtabular}
\end{table*}

\section{Remarks on multiple "interacting" current sheets.}
From \citealt{2018PhPl...25c2113P} we know that for a \emph{double} current sheet with inter-sheet half-distance $y_s$ and thickness $a$, the linear eigenmodes of the tearing instability are affected by the presence of other current sheets despite an inter-sheet distance $y_s/a\gg1$. Although the analysis in \citealt{2018PhPl...25c2113P} is valid in a resistive fluid approach, mass ratio and specific equilibrium prescription, it can be generalized for collisionless plasmas and different equilibrium configurations, see e.g. \citealt{DelSarto2016} for a proton-electron plasma.
We then expect, from the linear phase of the tearing instability --if present--, the growth of coupled modes between the multiple current sheets, i.e. the interaction occurs from time $t=0$. Still, possible competitor instabilities due to the proximity of the current sheets -- like the kink instability -- are expected to develop at later times ($\tau_{MHD}\sim L/v_A$), allowing tearing modes, developing here on kinetic scales ($\tau_{kin}\sim \omega^{-1}_{ci}$), to kick in first.
We will then improperly refer to the ``current sheets interaction'' discussing the visible non linear interplay between plasmoids developed originally in different current sheets.
We emphasize that for example in an extended striped-wind configuration the number of current sheets is in principle very large\cite{Sironi2011} and, as mentioned above, we expect a significant number of current sheets in the Crab flaring region as well. However, for a system with approximately uniform spacing between layers, the local reconnection physics is primarily controlled by the inter-sheet separation and the plasma parameters, rather than by the global number of sheets. Our periodic four-layer setup should therefore be viewed as a local “unit cell” of a much larger chain of current sheets, representative of a segment of a pulsar wind or other multi-layer systems.

\section{Simulation Results}
\begin{figure*}[t]
 \centering
 \includegraphics[width=0.8\linewidth]{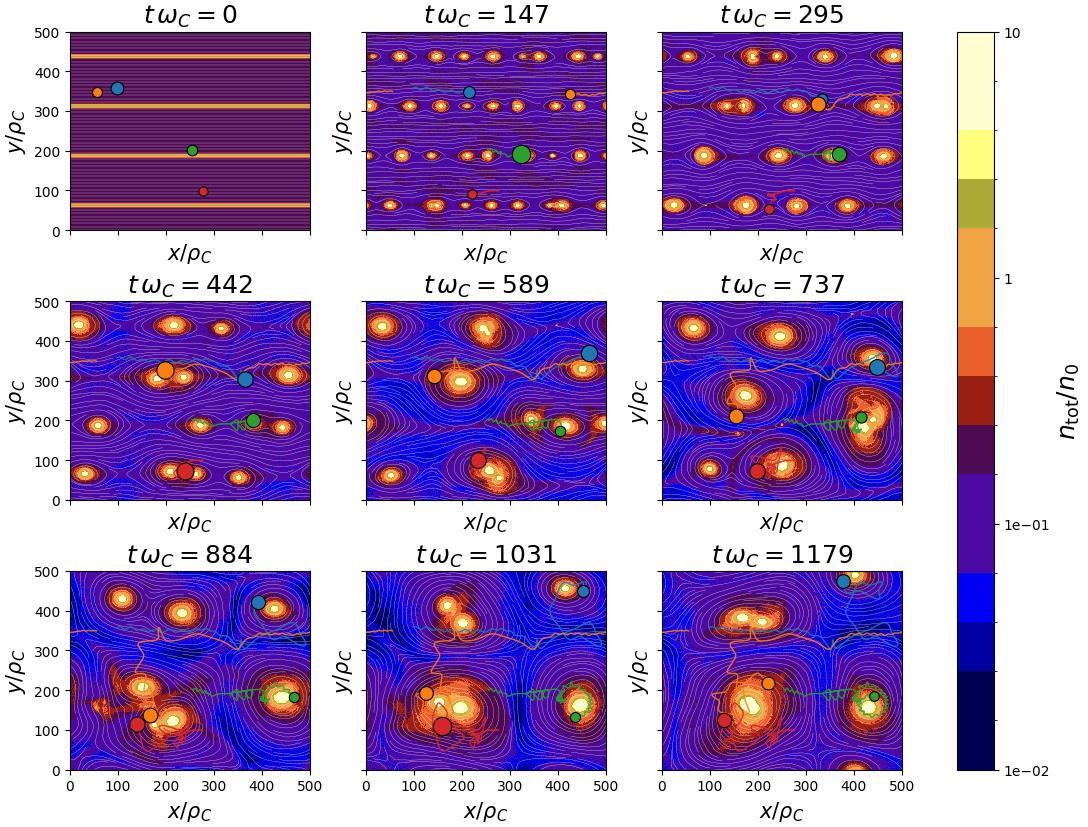}
 \caption{Color-coded plasma density at different times, normalized to the initial drifting particle density $n_0$. Superimposed magnetic flux function iso-contours. We also show some of the tracked particles, labelled with circles, whose size depends on the particle energy to show energy gains and losses. We also show the particle orbits to trace back the particle location at the energization region.}
  \label{fig:Density}
\end{figure*}

The simulation shows the development of the tearing instability, followed by the formation of plasmoids and a merging phase within the current sheets, similar to that observed in \citet{Cerutti2013}, see Fig.\ref{fig:Density}. After $t\omega_c\simeq 737$ macroscopic islands from different current sheets start merging. 

\subsection{Evolution and energy transfer to the plasma.}
\begin{figure}[t]
 \includegraphics[width=\linewidth]{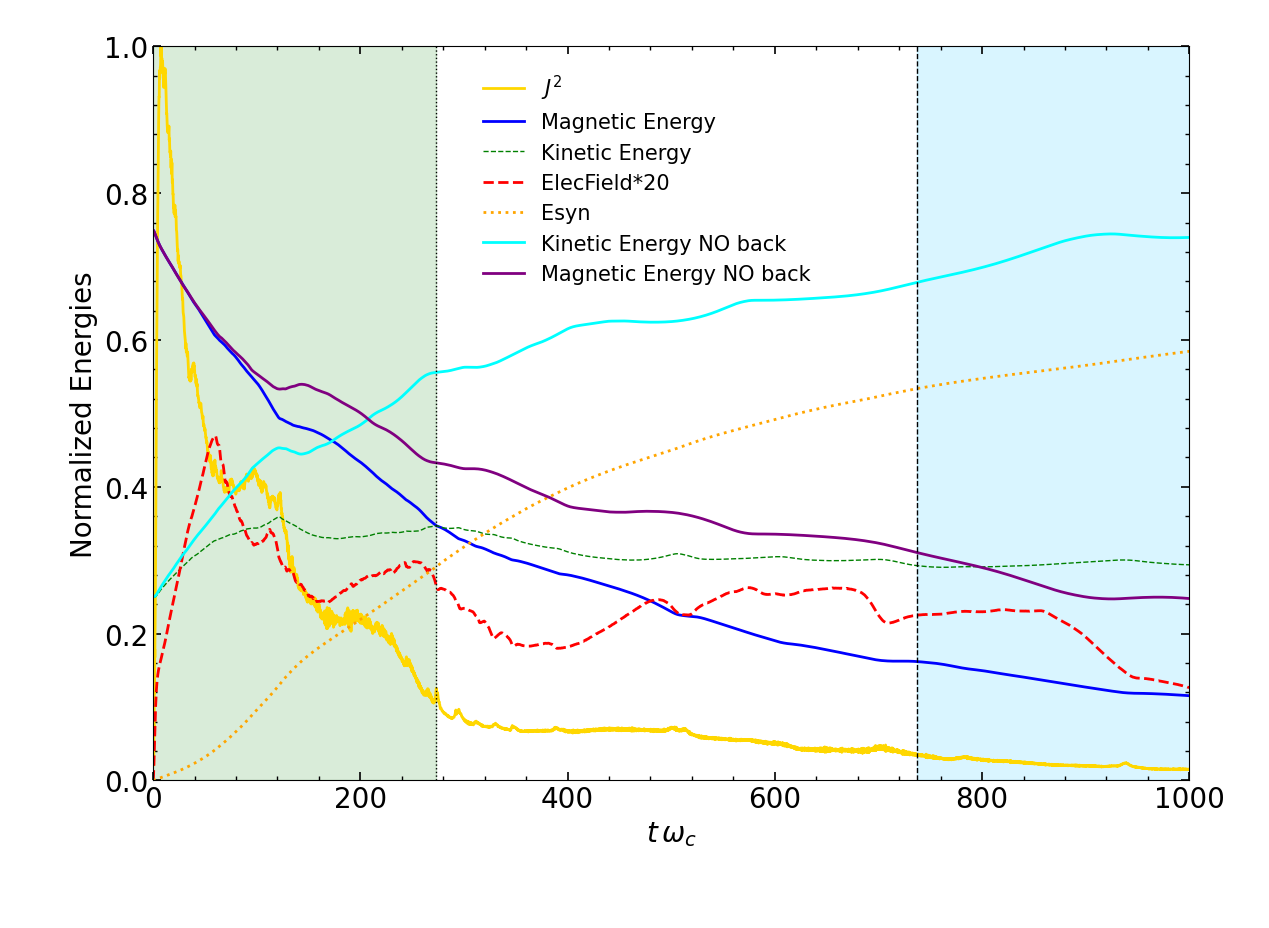}
 \caption{Energy normalized to the initial total energy available in the system. Magnetic energy with (blue) and without (purple) radiation back reaction. Electric field energies multiplied by a factor 20 for readability, with back reaction (red dashed line, similar to the case when radiation is off). Kinetic energy of all the particles (green dashed line with back reaction, cyan without). Orange dotted line marks the total energy lost through synchrotron cooling. In gold, the current dissipation $J^2$ rms, normalized to its peak value. Green shaded area marks the evolution before the equipartition between magnetic and kinetic energy for the simulation with back reaction, while the light blue shaded area marks the merging phase between plasmoids in different current sheets.}
  \label{fig:energy}
\end{figure}

Fig. \ref{fig:energy} shows the temporal evolution of the normalized energy components: magnetic energy dominates during the initial phase, gradually decreasing as it is converted into particle kinetic energy and radiation. 
In Fig. \ref{fig:energy} we also show the synchrotron cumulative emission from the accelerated particles, which increases monotonically over time as expected.
When synchrotron radiation is switched on, equipartition of magnetic and kinetic energy -- marked by the green shaded area --  occurs about $100\, \omega_c^{-1}$ later than the case without back reaction on particles.  
In both cases, kinetic energy rises rapidly and reaches a plateau, which sets at significantly lower energy in the case back reaction is switched on. 
To understand the energy transfer in Fig. \ref{fig:energy} we also plotted the quantity $J^2$ which is as a proxy for dissipation: it is associated with the steepening of magnetic-field gradients hence current concentration. A delay in the peak of the electric field with respect to $J^2$ suggests the current enhances during plasmoid compression, then triggering magnetic reconnection. After the kinetic energy and magnetic energy reach equipartition, the peaks of $J^2$ become hardly visible on the plot. Still, local enhancements are visible at $t\omega_c\simeq 500,700,950$, when we see in Fig. \ref{fig:Density} the merging phase between different current sheets is still ongoing.

\begin{figure}[t]
\includegraphics[width=1.0\linewidth]{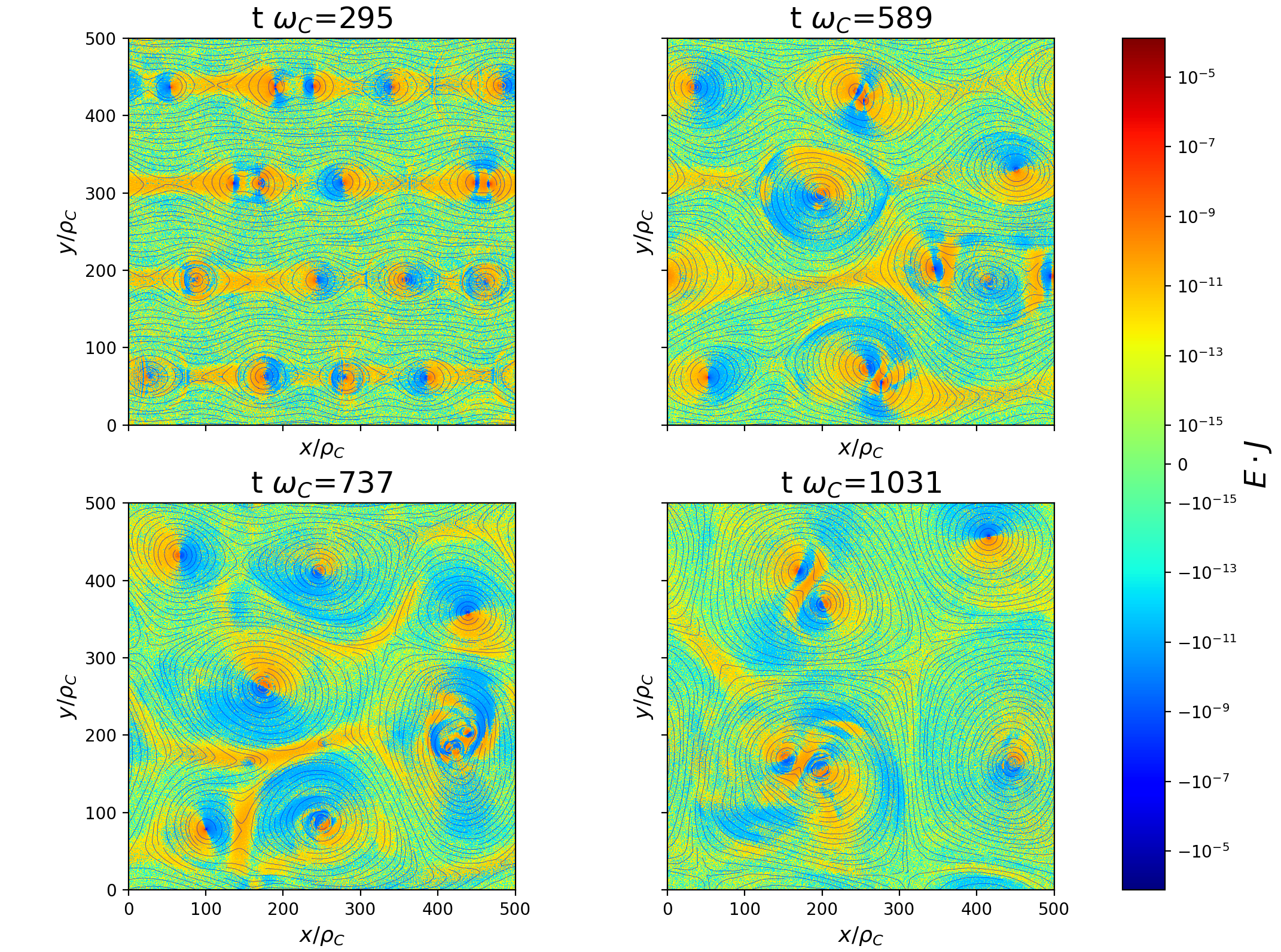}
 \caption{Energy transfer from magnetic field to particles ${\bf E}\cdot {\bf J}$, ${\bf J}$ being the total current carried by positrons and electrons. Superimposed (solid blue lines), magnetic field lines.}
  \label{fig:J24panels}
\end{figure}

The quantity ${\bf E}\cdot {\bf J}$, plotted in Fig.~\ref{fig:J24panels},
shows the energy transfer from fields to the plasma. The maxima are in correspondence of the current sheets and islands merging (see e.g. \citealt{2010ApJ...709..963D}) for primary as well as for secondary merging, confirming energy conversion at this locations by the reconnection electric field. While still a factor 20 smaller than the magnetic energy, the electric field is slightly larger than what found by \citealt{Cerutti2013}, probably due to the enhanced dynamics of the four coupled current layers.

\section{Turbulent Evolution}
In this section we examine the magnetic energy distribution from scales of about $100 d_e$ down to a few $d_e$ and compare magnetic spectra with reference slopes, to understand at which scales we can expect different physical mechanisms responsible for the cascade and dissipation of fluctuations in the system.  

\subsection{Fourier Analysis and Isotropic Spectrum}

To characterize the spatial distribution of magnetic fluctuations we compute
the two–dimensional Fourier transform of each field component 
$B_x(x,y)$, $B_y(x,y)$ and $B_z(x,y)$ over the simulation domain. 
We define the discrete Fourier transform as
\begin{align}
\tilde{B}(k_x,k_y) = 
\sum_{n=0}^{N_x-1}\sum_{m=0}^{N_y-1}
B(x_n,y_m)\,
e^{-i(k_x x_n + k_y y_m)},
\end{align}
where the discrete wavenumbers are
\begin{align}
k_x = \frac{2\pi}{L_x}\, m_x, \qquad
k_y = \frac{2\pi}{L_y}\, m_y.
\end{align}
\medskip
\noindent
The spectral power associated with each component of the magnetic field field is calculated as
\begin{align}
P_J(k_x,k_y) = \left|\,\tilde{B}(k_x,k_y)\,\right|^2 .
\end{align}
We then construct the isotropic 1D spectrum $E(k)$ by integrating the 
2D power over circular shells in $(k_x,k_y)$–space, 
with $k = \sqrt{k_x^2 + k_y^2}$ being
the radial distance in mode space.  
In this work we compute the isotropic spectrum summing over rings, which preserves the total
spectral power when integrated over all shells.

\medskip
\noindent
The isotropic spectra $E_{Bx}(k)$, $E_{By}(k)$ and $E_{Bz}(k)$ of the three magnetic
components are computed independently, and the total magnetic spectrum is
obtained as
\begin{align}
E_{B,\rm tot}(k) = E_{Bx}(k) + E_{By}(k) + E_{Bz}(k).
\end{align}

\begin{figure*}[t]
 \centering
 \includegraphics[width=0.9\linewidth]{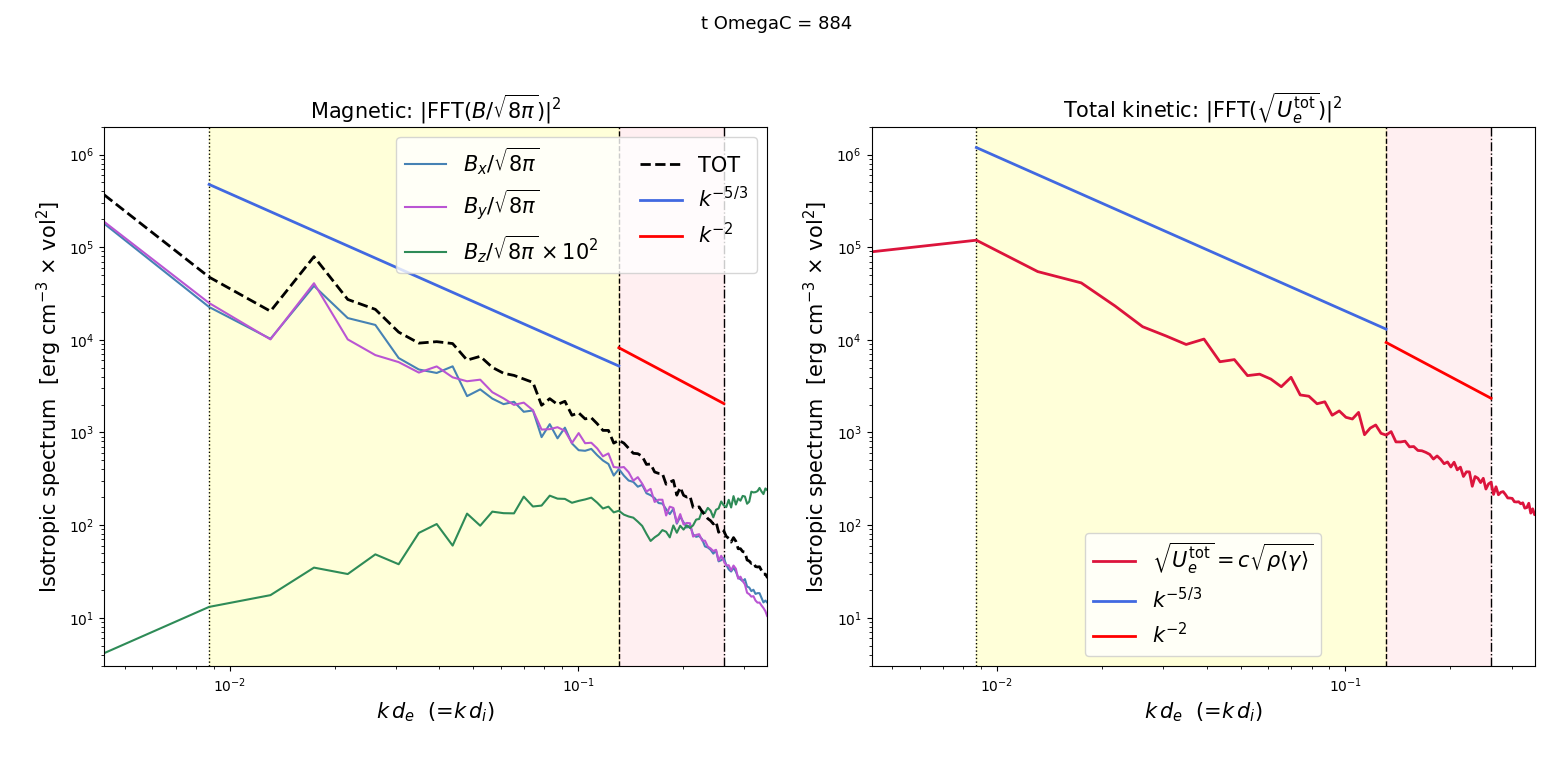}
 \caption{(Left)Isotropic 1D spectra of magnetic energy components $E_{Bx}(k)$, $E_{By}(k)$, $E_{Bz}(k)$ and total magnetic energy $E_{B,\rm tot}(k)$ at $t\omega_c=884$, together with the kinetic power spectrum $E_K(k)$} (see text for the detailed calculation).
  \label{fig:spectrum}
\end{figure*}

In Fig. \ref{fig:spectrum} (left) we show the results at $t\omega_c=884$--similar results we obtain a slightly later and earlier times. The yellow shaded area shows a scaling compatible with $k^{-5/3}$, the hallmark of an inertial-range cascade of magnetized Kolmogorov turbulence
\citep{Bruno2016}. Even if the slope extends only over about a decade, due to the limited resolution of the simulation, the range would appear to indicate the existence of a self-similar, conservative transfer of energy from large to smaller scales. Note that there is currently a debate on how reconnection dynamics affects the turbulent cascade in MHD, see, e.g., the review by Schekochihin\citep{Scheko22}. 
Nonetheless MHD turbulent phenomenologies all yield magnetic field spectra with negative power-law exponents in the range $3/2$\,---\,$5/3$. Here the nonlinear interactions proceed via reconnection and merging of plasmoids, originating within but also between different current sheets. 

At higher wavenumbers $k\,d_e>2\times 10^{-1}$, which in terms of length-scales translates in $l\sim 5d_e$, the steepening transitions toward a $k^{-2}$ spectrum, which is indicative of a dissipation range. This break, usually expected at $l\sim d_e$, is consistent with theoretical predictions from kinetic turbulence \citep{Scheko22, Bruno2016} and with observations in 
collisionless plasmas (e.g. in spacecraft measurements where the magnetic-field spectrum transitions from an 
MHD-like inertial range to a kinetic dissipation range, with slopes\citep{Leamon1998, Smith2006, Alexandrova2009, Sahraoui2009, Chen2013} between $-2$ and $-4$)

Since the system reaches equipartition before the turbulent phase, see Fig. \ref{fig:energy}, we also computed the kinetic energy spectrum.
In a relativistic plasma, when $\gamma \gg 1$ as in our case, the total kinetic energy density (bulk + thermal)
is obtained by summing over all particles $s$:
\begin{equation}
    U_e^{\rm tot} = \sum_s \gamma_s\, m_s c^2\, n_s,
\end{equation}
Our system is compressible, hence in analogy with the kinetic energy plotted in see Fig. \ref{fig:energy}, we define
\begin{equation}
    \mathcal{K} \equiv \sqrt{U_e^{\rm tot}} 
    \qquad [\mathcal{K}] = \text{erg}^{1/2}\,\text{cm}^{-3/2},
\end{equation}
whose square $\mathcal{K}^2 = U_e^{\rm tot}$ is the total kinetic energy density.
The kinetic power spectrum is then $E_K(k) = \sum_{|\mathbf{k}'|\approx k}|\tilde{\mathcal{K}}(\mathbf{k}')|^2$.
If $E_K(k) \approx E_B(k)$ at a given scale $k$, the system is locally
in kinetic--magnetic equipartition at that scale.
We stress that $U_e^{\rm tot}$ includes \emph{both} the bulk kinetic energy
and the thermal (internal) energy of the particles, since for each particle
$E_s = \gamma_s\, m_s c^2\, n_s$ contains contributions from both ordered bulk motion
and random thermal motion.
In our simulations, the particles are ultra-relativistic with
$\langle\gamma\rangle \sim 10^7$, so the thermal energy strongly dominates
over the bulk kinetic energy $\tfrac{1}{2}\rho v_{\rm bulk}^2$.
The quantity $\sqrt{U_e^{\rm tot}}$ therefore provides a more complete
and physically meaningful measure of the kinetic energy content of the plasma
than the bulk velocity alone. The resulting spectrum is again in agreement with the $-5/3$ Kolmogorov spectrum. While \citealt{Kritsuk_2007} found that highly compressible turbulence leads to a different spectral scaling, our case likely involves less extreme density fluctuations. The behavior we found could also be explained with a cascade dominated by Alfvénic modes or in the presence of significant correlations between density and velocity fluctuations. A detailed investigation of these possibilities is left for future work.

\subsection{Particle acceleration and radiative emission.}

\begin{figure}[t]
  \centering
  \subfigure
  {
    \includegraphics[width=0.47\textwidth]{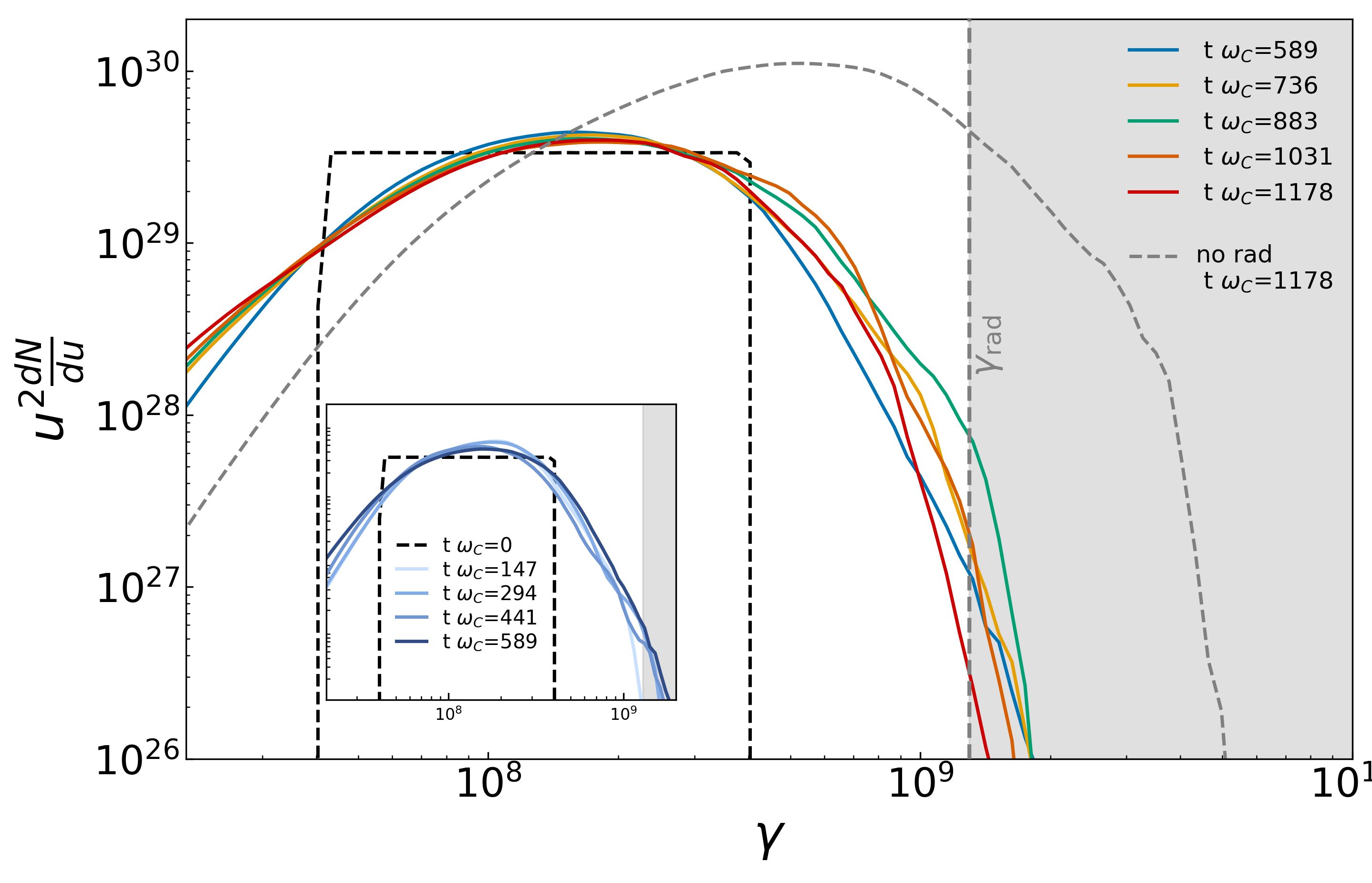}
  }\hfill
  \subfigure
  {
    \includegraphics[width=0.5\textwidth]{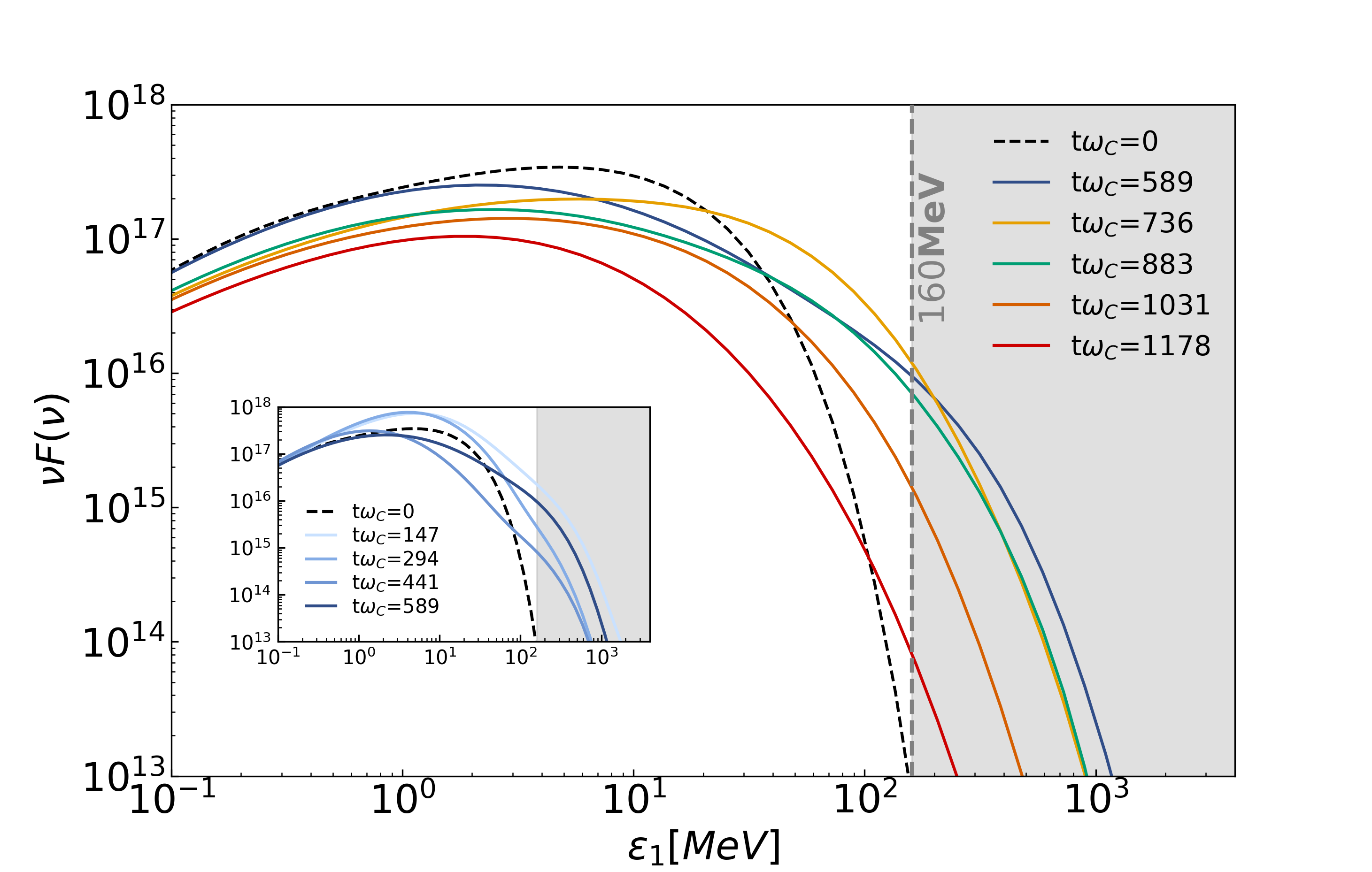}
  }
  \caption{Top: normalized background particle distributions $u^2N(u)$, initially a power-law with index $p=2$, (corresponding to a flat top hat with the normalization) at different times. Shaded gray area marks the value of $\gamma$ corresponding to synchrotron energy emission above the burn-off limit. Bottom: chromatic flux $\nu F(\nu)$, shaded gray area marks the energy above the burn-off limit. In both panels the insets show earlier time, prior to the merging.}
  \label{fig:twocol_subfigs}
\end{figure}

In Fig. \ref{fig:twocol_subfigs} (top panel) we show the particle distribution at different times: the inset shows the phase before merging while the main plot shows times after the current sheets started interacting. 
We can see the formation of a non-thermal tail beyond $\gamma= 10^9$, up to times $t\omega_c>1000$, well after the beginning of the merging phase. 
With $p$ defined as the background distribution power-law index at $t=0$, we indicate the background distribution power-law index at $t>0$ with $p_1$. In Fig. \ref{fig:twocol_subfigs} (top panel) a hard spectral index $p_1\gg 2$ is visible, especially in the initial phase, with a small number of particle populating the very high energy part of the spectrum. This differs with respect to the two current sheet case studied by \citealt{Cerutti2013}, in which at $t\omega_c\sim 400$ a non-thermal particle population with a slope $p_1\sim 2$ is observed. There are two possible explanations: (i) when the current layers start interacting with each other, the reconnecting field is not the initial magnetic field $B_0$ and since $\gamma_{rad}\propto B_0^{-1/2}$ the distribution new cut-off is determined by this new magnetic field value (ii) plasmoid merging makes the kinking of current sheets a favorable instability with respect to tearing, determining a different energy transfer from field to the plasma. In the case of four current sheets distributed over a domain with a double size along the y-direction (meaning the inter-distance between the current sheets is doubled) we expect the same behavior of the particle distribution functions as in \citealt{Cerutti2013}.
During the merging phase the high energy distribution displays a $p_1\gg 2$ at all the times sampled. We still can't exclude the presence of a short transient phase during which $p_1\sim 2$, despite in \citealt{Cerutti2013} this appears as a short, transient feature. Comparing to what found by \citealt{Nakanotani2022} in their MHD 2D simulation for the turbulent phase, we also found a steeper slope even in the case where we switched of the radiation reaction, as shown in Fig. \ref{fig:twocol_subfigs} (top panel, gray dashed line).This suggests the energization mechanisms in the MHD and kinetic framework might differ. 
Finally we observe that the low energy part of the spectrum is also populated, which is typical of diffusion mechanisms (see e.g. \citealt{Sironi2011}).
In Fig. \ref{fig:twocol_subfigs} (bottom panel) we show the chromatic flux $\nu F(\nu)$, with times prior to plasmoid merging in the inset.  
After the first development of plasmoids leading to emitted fluxes around $\epsilon_1=2\, 10^3 MeV$ the main plasmoid merging events (dark blue, green and gold curve) show emission up to about $\epsilon_1\le10^3 MeV$, while even for later times the flux at energies $\epsilon_1 \ge 160$ MeV remains non-negligible. 

\section{Orbits.}
We describe here some representative particle orbits showing acceleration during the simulations, and their radiative signatures. The sample of our tracked particle is a small subset of the total particles and just a few particles are accelerated to $\gamma>10^6$, hence the maximum synchrotron emission we found for this subset is less than the burn-off limit energy reference. Still, here the goal is to identify where and which process is providing energy to the plasma.

\begin{figure*}[t]
 \centering
 \includegraphics[width=0.8\linewidth]{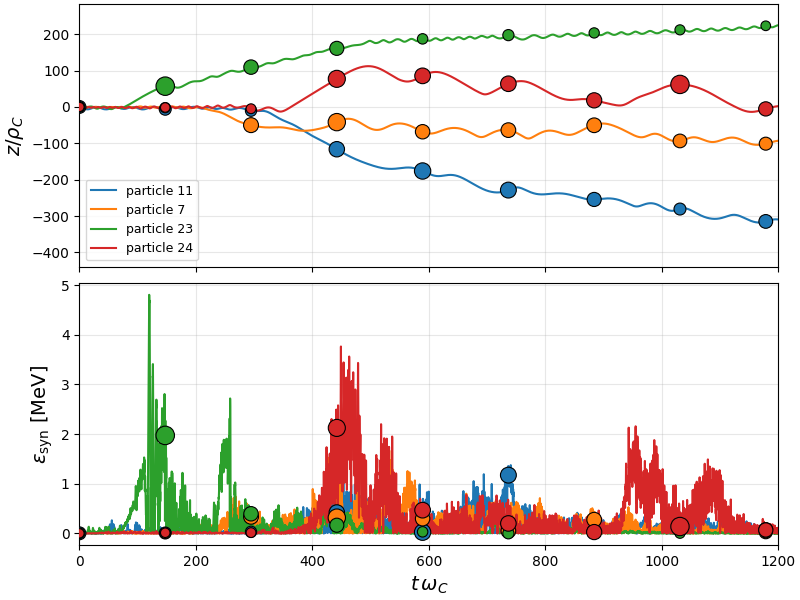}
 \caption{Top panel: trajectories (projected onto the yz-plane) of high-energy particles labelled in Fig.1. Bottom panel: critical synchrotron photon energy, see text for the definition.}
  \label{fig:orbits}
\end{figure*}

Fig.\ref{fig:Density} shows the trajectories in the $(x,y)$ and plane of particles labelled by different colored circles. The size of the circle represents the energy associated with the particle -- basically the particle $\gamma$. Before multiple current sheet merging at $t\omega_C=147$ the particle labelled in green gains significant energy in the the third from the top layer. In Fig. \ref{fig:orbits} we see the particle synchrotron critical energy $\epsilon_{sync}= 3heB\perp \gamma^2/4\pi mc$ spikes. At $t\omega_C=295$ the particle has entered an island and radiated most of its energy without any significant gain within the plasmoid. 
The particle labelled in orange, gains energy when entering the second from the top current sheet at around $t\omega_C=295$, then further accelerates at a plasmoid merging location (within the same current sheet) at $t\omega_C=442$. After the merging is over, at $t\omega_C=589$ we can see (orange solid line, tracking the particle orbit) that the orange particle reverses its motion, while it was initially bounded to be ejected from the current sheet in between the two merging plasmoids. The particle is now trapped in the resulting merged island. During the rest of the simulation it will remain within large plasmoids, radiating through multiple bursts with relatively low energy ($\epsilon \le 1$ MeV), see Fig. \ref{fig:orbits}.
Finally, the particle labeled with a red circle is the most representative of the merging dynamics. It gains energy around $t\omega_C=442$, subsequently radiating most of its energy, see Fig. \ref{fig:orbits} lower panel. After $t\omega_C=884$ it undergoes a secondary acceleration, reflected in Fig.\ref{fig:Density} in the particle marker size increase. This secondary energization is due to the crossing of the plasmoid merging region in the lower left quadrant around $t\omega_C=884$. In Fig. \ref{fig:orbits} we can see that at $t\omega_C>1000$ most of the particle energy is radiated as the particle is trapped in the newly formed big plasmoid in the left lower corner. Comparing Figs. \ref{fig:J24panels},\ref{fig:Density} we can see that the area the particle just crossed is characterized by high values of ${\bf E}\cdot {\bf J}$, suggesting again that energization is mostly due to the reconnecting electric field rather than Fermi acceleration within islands coalescing islands. 
We can then conclude most of the energy gains for the sampled particles occur through the reconnecting electric field. In  reconnecting systems first-order Fermi acceleration arises when particles undergo repeated reflections from contracting islands or from curved and relaxing magnetic field lines, hence gaining energy over multiple recurring reconnection cycles \citep{Drake2006,Dahlin2014}.
Here we expected turbulence to facilitate first-order Fermi processes by enhancing particle transport, rather than observing stochastic (second-order) Fermi acceleration. In particular, we were expecting turbulence to enable repeated encounters with current sheets and transport necessary for multiple acceleration to occur at the reconnection region. It is possible, however, that the turbulence developed here is not sufficient to produce persistent contraction or efficient/multiple return of particles to reconnection regions; in that case, the corresponding first-order Fermi mechanism is expected to remain  relatively inefficient. This picture can substantially change in three-dimensional simulations, where stochastic field-line wandering and enhanced transport allow energetic particles to escape local trapping and repeatedly access active acceleration regions \citep{Dahlin2017,Kowal2012,ComissoSironi2018}.

\section{Conclusions}

We have investigated the nonlinear evolution of multiple interacting current sheets under relativistic, radiative conditions using the \textsc{Zeltron} PIC code. 
Starting from a four-layer Harris equilibrium with alternating magnetic polarity, the system develops tearing instabilities on each sheet, followed by plasmoid formation and successive island merging. 
As reconnection proceeds, the interaction between adjacent sheets triggers kinking of these structures and a transition toward a turbulent regime characterized by enhanced electric fields and distributed dissipation sites. 
\begin{itemize}
\item The initial single current sheet acceleration phase is comparable to what found by \citealt{Cerutti2013}. Some differences emerge, due to the closer proximity of the current sheet configuration we adopted: (i)the highly dynamical multiple current sheet configuration provides slightly larger values of electric field; (ii) particle energy distribution never reaches a $-2$ slope, even if this could be a transient feature, as shown in \citealt{Cerutti2013}. The latter is most probably due to the redistribution of magnetic energy in shorter current sheets, subject to the islands dynamics, making particle acceleration at high energy dependent on the local magnetic field configuration. 
\item The evolution of the 4-sheet system evolves via magnetic reconnection and nonlinear interactions which lead to a fully developed turbulence at intermediate times, characterized by a magnetic energy spectrum with a slope that is comparable to a Kolmogorov $-5/3$ scaling. The spectrum, well-developed for just over a decade, then steepens at kinetic scales where energy is dissipated. Much higher resolution simulations would be required to
glean the details of spectrum formation and interplay between reconnection and cascade in the formation of the spectral slope.
\item In this turbulent phase, comparing to what found by \citealt{Nakanotani2022} in their MHD 2D simulation, we didn't find the same $-2$ spectral slope in the turbulent regime. This is also valid in the case the back reaction is not switched on. This suggests the energization mechanisms in the MHD and kinetic framework might differ, but further simulations should be run to assess the role of a sheared field with respect to the pressure equilibrium adopted in this work. Our results for the kinetic energy spectrum suggest a scenario with more moderate density fluctuations with respect to \citealt{Kritsuk_2007}, and are consistent with a Kolmogorov-like scaling. This may reflect a cascade dominated by Alfvénic dynamics or the presence of correlations between density and velocity fluctuations, the exploration of which is deferred to future works.
\item The synchrotron radiative energy losses act as an active feedback mechanism at all times, especially during the initial turbulent phase: magnetic energy is mainly converted in radiation loss. In the case without any radiation reaction all of the energy is instead converted into kinetic energy of the plasma.
\item Particles tracked show acceleration occurs mainly in reconnecting current sheets, while no major acceleration by the Fermi mechanism is shown in particles trapped in plasmoids.
\end{itemize}
Overall, our results highlight that the inclusion of multiple current sheets within the same 'emission region' substantially modifies the morphology and energetics of reconnection: inter-layer coupling accelerates the onset of turbulence, enhances radiative efficiency, and leads to a more homogeneous dissipation of magnetic energy. 
These findings provide a framework for understanding the role of magnetic reconnection in episodic high-energy flares observed in the Crab Nebula, where multi-layer current sheets may be expected. The theoretical analysis carried out here may be applicable, with an appropriate change in parameters, to many other environments: for example, to the fate of the pulsar striped wind.

\begin{acknowledgments}
Simulations were carried out with the PLEIADI (Catania) supercomputer by USC VIII-Computing of INAF. FP sincerely thanks Fabio Vitello and Salvatore Scavo for technical support. FP also thanks Eloisa Menegoni for discussions on reconnection simulations with Zeltron. 
EA and NB acknowledge support by the European Union —  NextGenerationEU RRF M4C2 1.1 under grant PRIN-MUR2022TJW4EJ. FP, MEI and KS acknowledges support from the Deutsche Forschungsgemeinschaft (DFG, German Research Foundation), within the Collaborative Research Center SFB1491 and grant number 544893192.
\end{acknowledgments}

\bibliographystyle{aipnum4-1} 
\bibliography{references}      

\end{document}